\documentclass{article}
\usepackage{spconf}
\usepackage{graphicx}
\usepackage{amsmath}
\usepackage{array}
\usepackage{amsthm,amssymb,fixmath,mathtools,nicefrac}
\usepackage{subfig,xspace,xcolor,tabulary}
\usepackage{float}
\newcolumntype{N}{>{\centering\arraybackslash}m{.5in}}
\newcolumntype{G}{>{\centering\arraybackslash}m{2in}}

\title{Segmentation-based Method combined with Dynamic Programming for Brain Midline Delineation}
\name{Shen Wang$^{1*}$, Kongming Liang\thanks{$^{*}$Equal contribution \protect\\
\indent\indent This work was supported in part by the following grants MOST-2018AAA0102004, BNSF-7191003, NSFC-61625201, NSFC-81830057.
}$^{3*}$, Chengwei Pan$^{2}$, Chuyang Ye$^{6}$, Xiuli Li$^{2}$, Feng Liu$^{2}$,Yizhou Yu$^{2}$, Yizhou Wang$^{3,4,5}$}
\address{$^1$Center for Data Science, Peking University, Beijing, China;
$^2$Deepwise AI Lab, Beijing, China \\
$^3$Computer Science Department, School of EECS, Peking University, Beijing, China\\
$^4$Center on Frontiers of Computing Studies, Peking University, Beijing, China \\
$^5$Advanced Institute of Information Technology, Peking University, Beijing, China \\
$^6$School of Information and Electronics, Beijing Institute of Technology, Beijing, China}

\begin{document}
%
\maketitle
\begin{abstract}
The midline related pathological image features are crucial for evaluating the severity of brain compression caused by stroke or traumatic brain injury (TBI). The automated midline delineation not only improves the assessment and clinical decision making for patients with stroke symptoms or head trauma but also reduces the time of diagnosis. Nevertheless, most of the previous methods model the midline by localizing the anatomical points, which are hard to detect or even missing in severe cases. In this paper, we formulate the brain midline delineation as a segmentation task and propose a three-stage framework. The proposed framework firstly aligns an input CT image into the standard space. Then, the aligned image is processed by a midline detection network (MD-Net) integrated with the CoordConv Layer and Cascade AtrousCconv Module to obtain the probability map. Finally, we formulate the optimal midline selection as a pathfinding problem to solve the problem of the discontinuity of midline delineation. Experimental results show that our proposed framework can achieve superior performance on one in-house dataset and one public dataset. 
\end{abstract}
\section{Introduction}
\label{sec:intro}
\begin{figure}[t]
\centering
\includegraphics[scale=0.45]{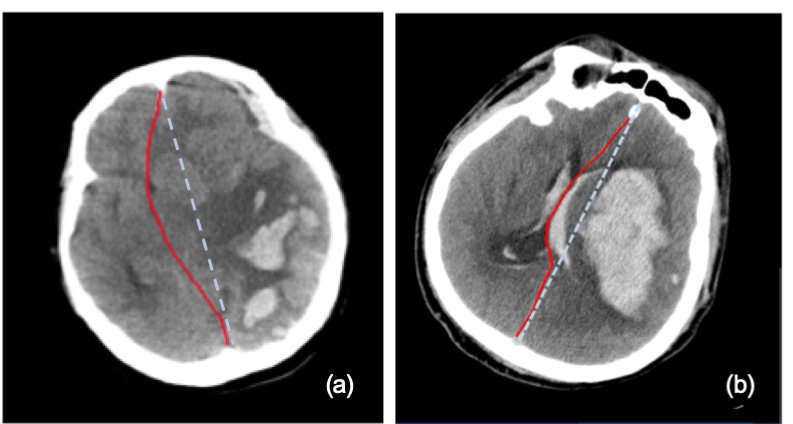}
\vskip 2mm
\caption{Examples of ideal midlines (light blue dotted line) and actual midlines (red line). (a) and (b) are two examples of severe midline shift.} \label{fig_task}
\vskip -2mm
\end{figure}

The anatomical structure of the human brain consists of two symmetrical hemispheres, which are separated by the ideal midline (as shown in Fig.~\ref{fig_task}). In general, midline structures in the non-contrast CT images are often associated with high intracranial pressure. Therefore, they provide abundant information for physicians to make an accurate diagnosis on the severity of stroke or TBI e.g. Immediate surgery may be indicated when there is a midline shift of over 5 mm~\cite{tu2012postoperative}.

Automated midline delineation can quantify midline shift and speed up brain interpretation and decision making of foremergency physicians. Since the brain CT reading reliability of the emergency physicians is often questioned~\cite{dolatabadi2013interpretation}, the quantification results of the midline can improve the assessment of stroke or TBI. Combined with other information (e.g. gender, age), the pathological image features related to midline can benefit the clinical diagnosis and prognosis treatment.

\begin{figure*}[t]
\centering
\includegraphics[scale=0.55]{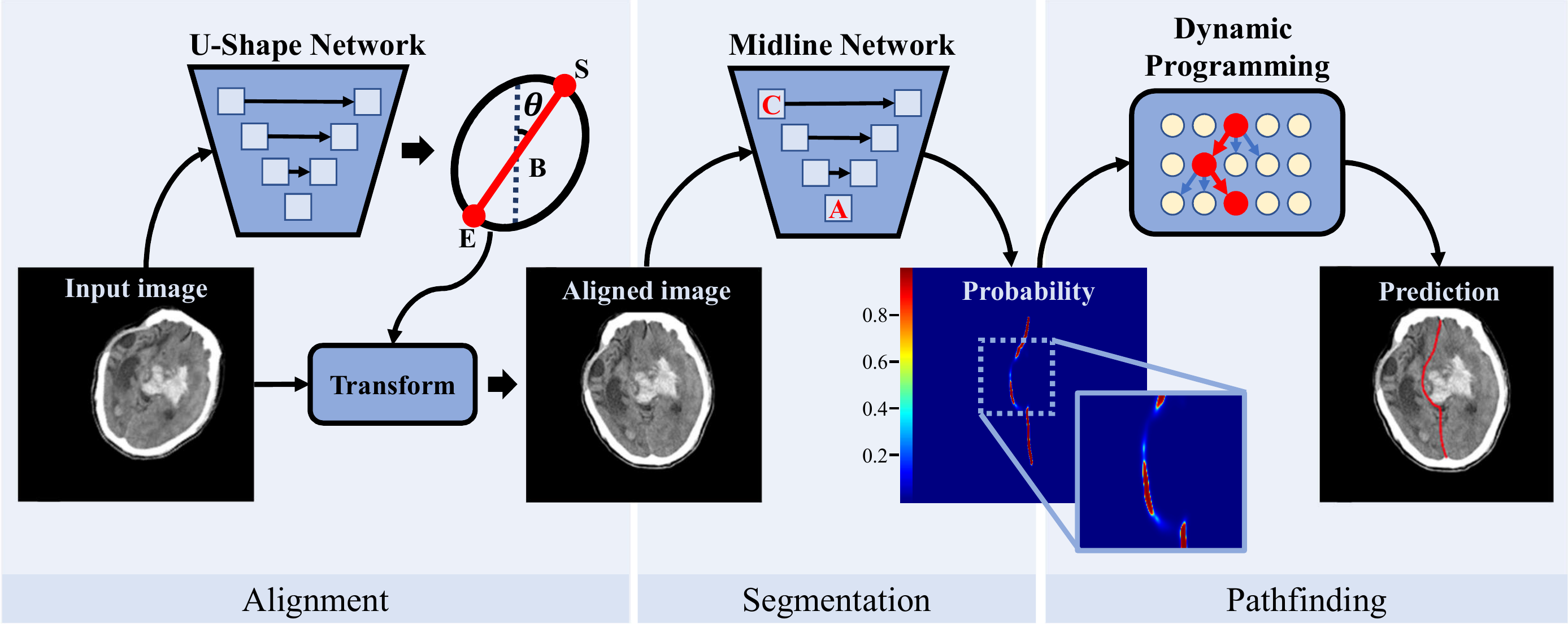}
\caption{The proposed framework consists of three stages. (a) Alignment: $S$, $E$, $B$ are start, end, center point of the predicted midline respectively and $\theta$ denotes the offset angle from the vertical direction; (b) Segmentation: C in the Midline Network denotes the CoordConv Layer and A denotes the Cascade AtrousConv Module; (c) Pathfinding.}  
\label{fig1}
\end{figure*}

Previous methods mainly focus on localizing the pre-defined points or parts based on anatomical information of the human brain. Liao \textit{et al.}~\cite{liao2006tracing} proposed a deformed midline model according to the biomechanical properties of intracranial tissue. Chen \textit{et al.}~\cite{chen2010actual} estimated the position of the midline using shape matching among multiple regions. Similarly, Qi \textit{et al.}~\cite{qi2013automated} presented a variational level set to extract the ventricle contours. Then the position of midline was detected based on the identified right and left lateral ventricle contours. Liu \textit{et al.}     \cite{liu2014automatic} proposed to delineate the midline by automatically localizing the anatomical points. However, in severe cases(see Fig~\ref{fig_task}), the predefined anatomical points or parts of the human brain may not be visible which limits the above methods. Besides, the detected midline referring to anatomical points is usually not smooth and decreases the quantification quality for an accurate diagnosis.

To address such issues, we formulate the brain midline delineation as a segmentation task and propose a novel three-stage framework. By modeling the midline in a pixel-wise way, our method can accurately quantify the pathology features which are essential for clinical applications. 
Besides, based on the UNet~\cite{ronneberger2015u}, we propose to inject a CoordConv Layer to leverage the spatial information of brain midline and a Cascade AtrousConv Module to enlarge the receptive field for the further refinement. Moreover, we formulate the optimal midline selection as a pathfinding problem to solve the problem of the brain midline discontinuity. This strategy can be applied to other segmentation tasks with geometric constraints. Experimental results show that our proposed framework achieves superior performance and promising generalization ability on automatic midline delineation.

\section{Method}

Our proposed framework is illustrated in Fig.~\ref{fig1}. It comprises three stages: image alignment, midline segmentation and pathfinding. In the following sections, we will introduce each stage in detail.


\subsection{Alignment}
\label{sec:align}
Due to the variations of patients' head positions during CT scanning, the relative position of the brain in the CT image is usually not consistent. As the midline delineation is sensitive to the location information, we attempt to align the image to the standard space. We first use a standard UNet\cite{ronneberger2015u} to estimate the initial midline structure. Consequently, the two endpoints (S and E in Fig. \ref{fig1}) of the midline can be located to calculate the offset angle and the brain center. We can align the original CT image by an affine transformation of translation and rotation so that the semantics of pixels in CT of different directions can be aligned.


\subsection{Midline Detection Network}
Our proposed midline detection network (MD-Net) is based on the UNet. We adopt the CoordConv Layer for spatial information modeling, which is set as the first layer of the encoder. Besides, a Cascade AtrousConv Module is added to the highest semantic level feature, which can enlarge the receptive field remarkably. 

\noindent \textbf{CoordConv Layer.} The midline is the junction of the two hemispheres, which is highly correlated to position information. In this part, we adopt the coordination-guided convolutional layers (CoordConv Layer)~\cite{liu2018intriguing} to model the spatial information. The CoordConv is a simple extension to the classic convolutional layer, which integrates position information by concatenating extra coordinate channels. Two extra channels are added respectively to represent x, y coordinates of the input. The values of coordination channels are normalized to the range from -1 to 1. 

\begin{figure}[t]
\vskip -3mm
\centering
\includegraphics[width=0.37\textwidth]{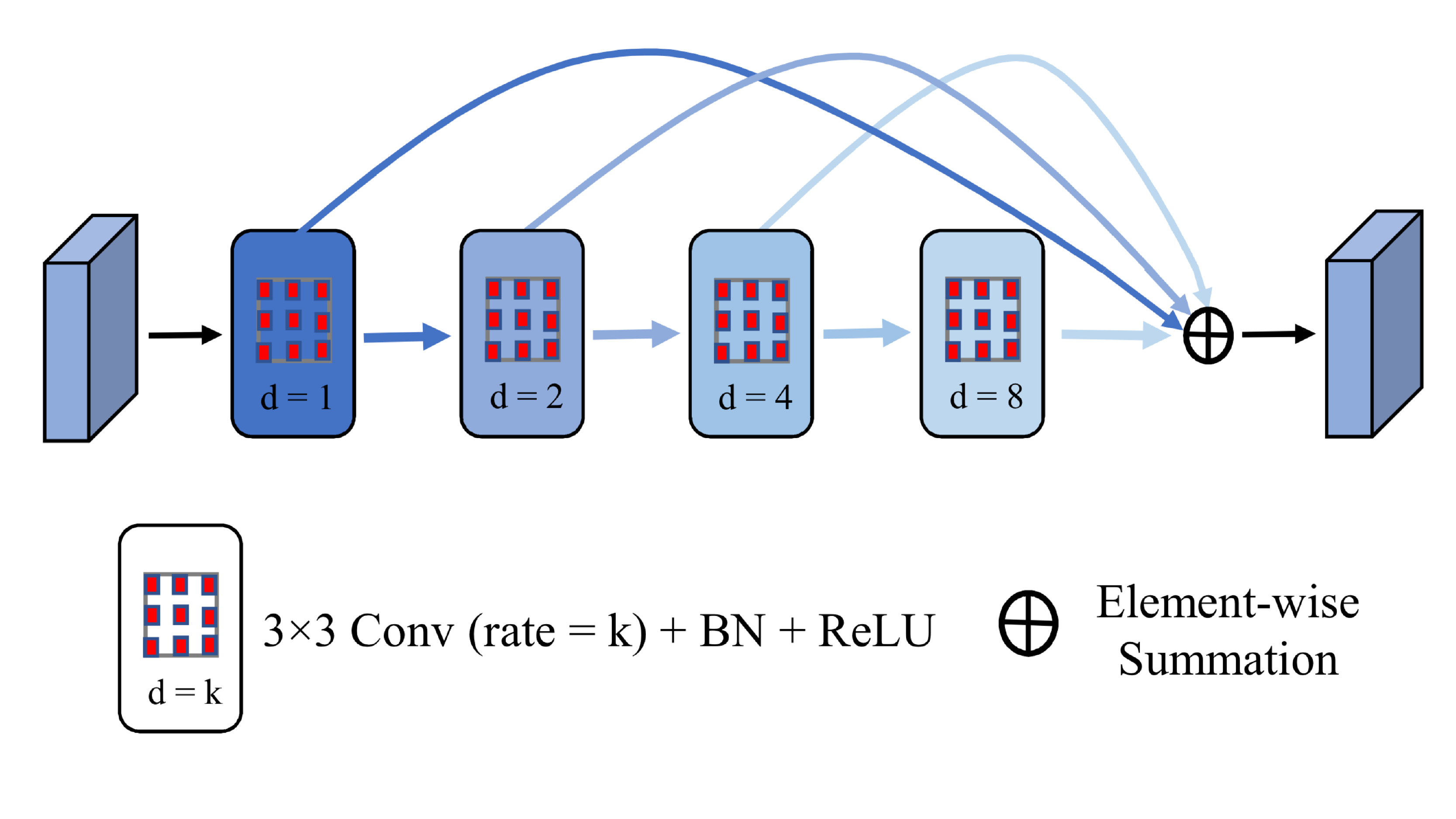}
\caption{The Cascade ArtousConv Module stacks dilated convolutions in cascade mode. The dilation rates of the stacked dilated convolution layers are 1, 2, 4, 8 respectively.
}  
\label{fig_network}
\vskip -3mm
\end{figure}

\noindent \textbf{Cascade AtrousConv Module.}
Atrous convolutions are widely used for semantic segmentation~\cite{chen2014semantic}, which increase the receptive field while keeping the feature map resolution unchanged. The receptive field of the encoder in the standard UNet is 140$\times$140, only covering part of the input CT image, which may neglect the importance of global context information on midline segmentation. 
Therefore, we propose a Cascade AtrousConv Module to explore a larger receptive field (620$\times$620, covering the full image 512$\times$512), as shown in Fig.~\ref{fig_network}.

\noindent \textbf{Loss Function.} Cross entropy is a classic loss function in semantic segmentation tasks. Dice coefficient loss can alleviate the problem of sample imbalance to a certain extent. So we combine the weighted cross-entropy and dice loss as the total loss function $\mathcal{L}_{total} = \mathcal{L}_{wce} + \mathcal{L}_{Dice}$. The deep supervision mechanism is further imposed on the semantic maps of different scale.




\subsection{Midline Pathfinding}
In the standard space, the midline is a continuous line and composed of one point in each row. However, through our experiments, we find that the direct output of the network can not guarantee the continuity of the midline (see Fig.~\ref{fig1}). To address such an issue, we formulate the optimal midline selection as a pathfinding problem. Given a segmented probability map, the objective is to minimize the following energy function

\begin{equation}
\label{eqn_energy}
  E(\boldsymbol{p}) = \sum\limits_{i} \psi_i(p_i) + \sum\limits_{i,j} \psi_{ij}(p_i, p_j)  
\end{equation}
where $p_i$ represents the selected pixel in $i$-th row. We use a unary potential $\psi_i(p_i)=-log P(p_i)$, where $P(p_i)$ is the label assignment  probability at pixel $p_i$ as computed by Midline Network. The pairwise potential $\psi_{ij}$ represents the smooth terms between selected points in the two rows. To simplify the problem, only smooth terms between adjacent rows are considered. And we limit the direction of connectivity to be one of \textit{down}, \textit{bottom-left} and \textit{bottom-right}(see Fig. 2), so that $\psi_{ij}(p_i,p_j)$ is described as:

\begin{equation}
 \psi_{ij}(p_i,p_j) = \left\{ \begin{split}
    &1, \qquad \qquad \left|x_{p_i}-x_{p_j}\right|\leq1, \left|i-j\right| = 1 \\ 
    & + \infty, \qquad  \ otherwise \\
  \end{split}
\right.
\end{equation}
where $x_{p_i}$ denotes the horizontal coordinate of pixel $p_i$.

As we limit the connected directions between adjacent rows, the optimization problem is equivalent to find a path from the start point selected in the first row to the end point selected in last row, minimizing the above energy function defined in Eqn.~(\ref{eqn_energy}), which can be solved by \textit{dynamic programming}. Our proposed midline pathfinding algorithm has two advantages. Firstly, it takes the global information into account, so that the prediction of points is interdependent. Secondly, the constraints we add can guarantee the continuity and smoothness of the predicted midline.

\begin{figure*}[t]
\centering
\includegraphics[scale=0.45]{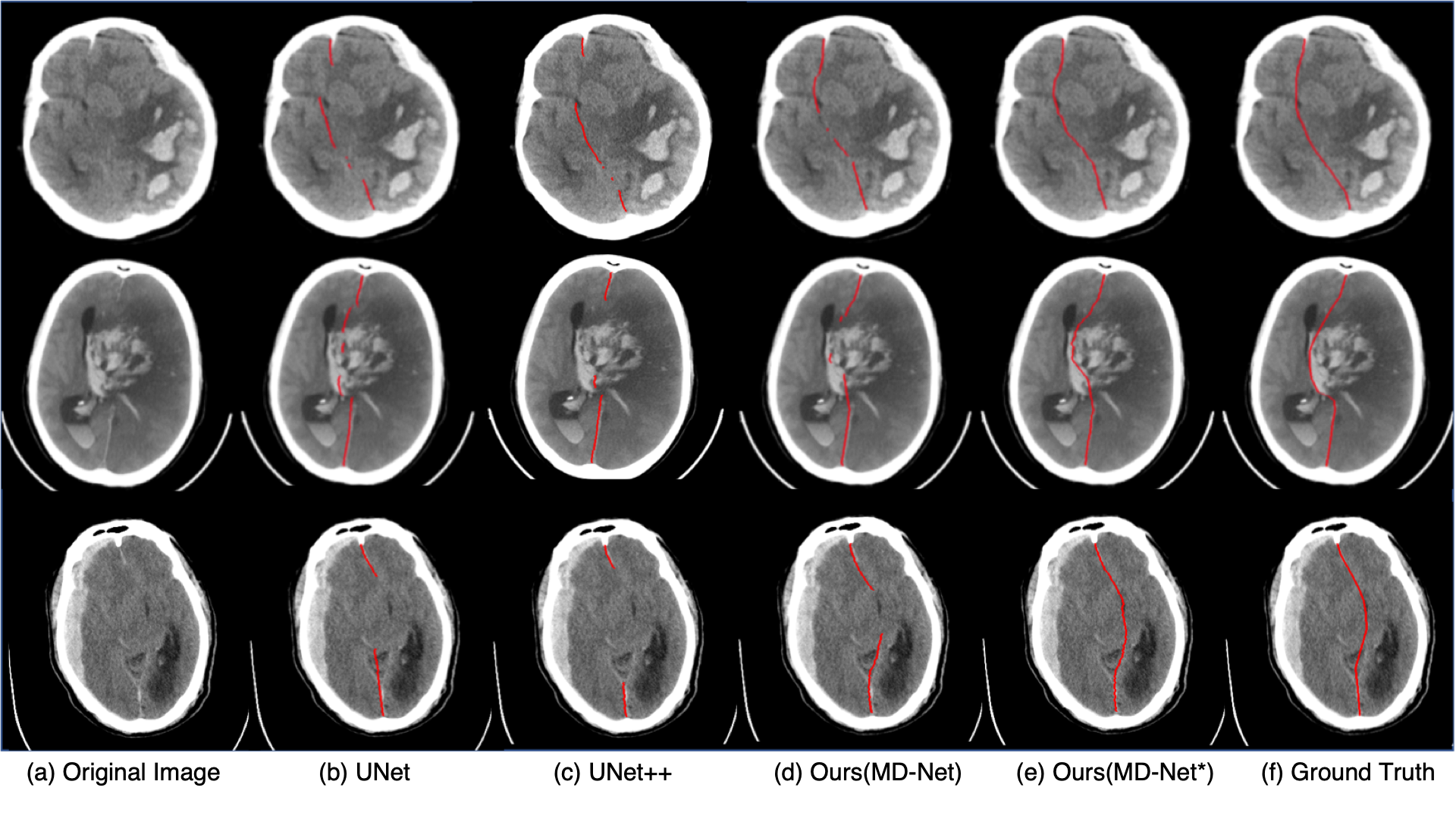}
\vskip -4mm
\caption{Qualitative comparison between UNet, UNet++ and proposed MD-Net, showing the results for the in-house dataset(the first row) and CQ500(the last two rows). '*' means the output probability map is post-processed during pathfinding stage.} \label{fig_result}
\vskip -3mm
\end{figure*}

\section{Experiments}

\subsection{Datasets and Evaluation Metrics}
\textbf{Datasets.} The proposed models are evaluated on one in-house dataset and one public dataset CQ500\footnote{http://headctstudy.qure.ai/dataset}. For the in-house dataset, we collected  877 non-contrast head CT scans with 5-mm slice thickness from three hospitals. The dataset is randomly split into a train/val/test set of 708/87/82 stacks and the number of scans with midline shift is 207/44/42 respectively. For the public dataset CQ500, we choose 235 CT scans (53 midline shift) with around 5-mm slice thickness. One senior radiologist marks all the CT scans as the gold standard.


\noindent \textbf{Evaluation Metrics.} 
The Dice coefficient, Hausdorff Distance(HD) and average symmetric surface distance(ASD) are the three most commonly used evaluation metrics in medical image segmentation~\cite{menze2014multimodal, yue2019cardiac}. In the midline delineation task, we need to measure the distance between the predicted midline and the actual midline which can not be well gauged by the Dice coefficient. Therefore, we choose the HD and ASD as the evaluation metrics. 


\subsection{Implementation details}
For image preprocessing, three adjacent slices are stacked as the input to model the inter-slice information and the image densities are normalized using the brain window. The midline ground truth is expanded to a band with 5-pixel width. 
We use Adam to train the model by setting  $\beta_1$ = 0.9, $\beta_2$ = 0.99 for 100 epochs. The initial learning rate is 0.001. The poly learning rate policy is employed where the initial learning rate is multiplied by $(1-\frac{iter}{total\_iter})^{power}$ with power = 0.9. Our implementation is based on 
Pytorch package.

\subsection{Results}

\begin{table}[h]
\caption{Comparison of the alignment and pathfinding stage in the proposed framework.}
\vspace{-0.5cm}
\renewcommand\tabcolsep{4.0pt} 
\small
\begin{center}
\begin{tabular}{c|cc|cc}
\hline
Method & Alignment & Pathfinding & HD & ASD\\
\hline
\hline
UNet &  & &   4.62(6.29) & 1.45(1.07)   \\
UNet & \checkmark & &  3.54(4.18) & 1.60(0.77)   \\
UNet & \checkmark & \checkmark &  \textbf{2.44(2.88)} & \textbf{0.83(0.68)}   \\
\hline
\end{tabular}
\end{center}
\vspace{-0.3cm}
\label{align}
\end{table}

\noindent \textbf{Ablation study.} In this section, we set the UNet~\cite{ronneberger2015u} as the baseline method, which is widely used in the medical image segmentation. The quantitative performance is in terms of mean $\pm$ std of HD (mm) and ASD (mm) index. We conduct ablation experiments to investigate the proposed three components: alignment, segmentation and pathfinding. All ablation experiments are evaluated on the in-house test set.

Firstly, we verify the effectiveness of the alignment and pathfinding stage. As shown in Table~\ref{align}, the alignment of input images can greatly improve the performance. In the pathfinding phase, the performance can be further improved by post-processing the probability map with dynamic programming. Secondly, we add each of the proposed modules to baseline independently.  Meanwhile, the input images are aligned. As shown in Table~\ref{model}, the modules we proposed can improve the performance to a certain extent. Combining all modules, our proposed MD-Net achieves the best results.

\begin{table}[h]  
\small
\centering  
\caption{Comparison of different model structures in the proposed network on the in-house dataset.}  
\label{model}
\begin{tabular}{l|cc}  
     \hline
        Method & HD &ASD \\   
       \hline
       \hline
       UNet&3.54(4.18) & 1.60(0.77) \\
       + CoordConv&  3.17(3.02) & 1.50(0.72) \\
       + Cascade AtrousConv& 3.07(3.58) &  1.52(0.72) \\
       Ours(MD-Net)& \textbf{2.93(1.68)} &  \textbf{1.49(0.65)} \\
       \hline
\end{tabular}
\end{table}

As shown in Fig.~\ref{fig_result}(b), baseline methods suffer two major issues to correctly delineate brain midline: 1) inaccurate position of the midline and 2) discontinuity of the midline. To address the first issue, we propose the MD-Net with the CoordConv layer and the Cascade AtrousConv module, which can enhance the position discrimination ability and guarantee that segmented midlines are around correct location (see Fig.~\ref{fig_result}(d)). For the second issue, we propose the pathfinding stage to guarantee the continuity of the midline which is essential in severe cases (see Fig.~\ref{fig_result}(e)). 

\noindent \textbf{Comparison to the State-of-the-Art Methods.}
The encoder-decoder architecture like U-Net \cite{ronneberger2015u} and UNet++~\cite{zhou2018unet++} has achieved state-of-the-arts in many medical image segmentation tasks. In this section, we compare the proposed MD-Net with the above two methods. As shown in Table~\ref{result}, our proposed method can achieve superior performance on both the in-house dataset and CQ500 dataset. Finally, our proposed method achieves an average HD of 2.26 (mm) and ASD of 0.81 (mm) in our in-house dataset and an average HD of 2.58 (mm) and ASD of 0.72 (mm) in our in-house dataset.

\begin{table}[h]  
\caption{Performance comparison of our method with other methods. '*' means the output probability map is post-processed during pathfinding stage. }  
\centering  
\small
\label{result}
\renewcommand\tabcolsep{4.0pt} 
\begin{tabular}{l|cc|cc}  
     \hline
        Method &\multicolumn{2}{c|}{In-house}&\multicolumn{2}{c}{CQ-500}\\   
        &HD&ASD&HD&ASD\\
       \hline
       \hline
       UNet~\cite{ronneberger2015u}&3.54(4.18)&1.60(0.77) & 3.26(3.26)& 1.67(0.66)\\
       UNet++\cite{zhou2018unet++}&3.46(3.86)&1.57(0.75)&3.59(4.82)& 1.68(0.76)\\
       Ours(MD-Net)& 2.93(1.68)&1.49(0.65) & 3.08(2.72)&1.58(0.63) \\
       \textbf{Ours(MD-Net*)}& \textbf{2.26(1.71)}& \textbf{0.81(0.57)} &  \textbf{2.58(2.54)} &  \textbf{0.72(0.53)}  \\
       \hline
   \end{tabular}
\vspace{-0.2cm}
\end{table}

\section{Conclusion}
We propose a novel framework of midline delineation on non-contrast head CT scans, which consists of three stages: alignment, segmentation and pathfinding. Firstly, we align an input CT image into the standard space. Secondly, the MD-Net is proposed to enhance the position discrimination ability. Thirdly, the output probability map is post-processed to guarantee the continuity of the midline by using dynamic programming. Finally, experimental results demonstrate the superior performance of proposed method on both the in-house dataset and the CQ500 dataset. 

\bibliographystyle{IEEEbib}
\bibliography{isbi}

\end{document}